\documentclass[onecolumn,aps,nofootinbib,a4paper]{revtex4}

\usepackage{geometry}
\usepackage[ngerman,british]{babel}
\usepackage[utf8]{inputenc}
\usepackage[T1]{fontenc}
\usepackage{lmodern} 
\usepackage[tracking=true]{microtype}

\usepackage{amsmath,amssymb}
\usepackage{amsmath}
\usepackage{amsthm}
\usepackage{mathtools}
\usepackage{dsfont}
\usepackage{bbold}

\usepackage[hyperfootnotes=false]{hyperref}

\newcommand{\ie}{i.\,e.\ }
\newcommand{\cf}{cf.\ }

\newcommand{\eg}{e.\,g.\ }

\newcommand{\re}{\operatorname{Re}}

\newcommand{\ket}[1]{\left|{#1}\right\rangle}

\newcommand{\ketbra}[2]{\left|{#1}\middle\rangle\middle\langle{#2}\right|}

\newcommand{\ew}[1]{\langle{#1}\rangle}

\newcommand{\abs}[1]{\lvert#1\rvert}

\newcommand{\der}[2]{\frac{d^{#2}}{d {#1}^{#2}}}
\newcommand{\tder}[2]{\tfrac{d^{#2}}{d {#1}^{#2}}}
\newcommand{\pder}[2]{\frac{\partial^{#2}}{\partial {#1}^{#2}}}
\newcommand{\tpder}[2]{\tfrac{\partial^{#2}}{\partial {#1}^{#2}}}

\newcommand{\bx}{\mathbf{x}}

\newcommand{\bfr}{\boldsymbol{\mathfrak{r}}}
\newcommand{\bfR}{\boldsymbol{\mathfrak{R}}}
\newcommand{\dbfR}{\dot{\bfR}}

\newcommand{\br}{\mathbf{r}}
\newcommand{\dbr}{\dot{\mathbf{r}}}
\newcommand{\bR}{\mathbf{R}}
\newcommand{\dbR}{\dot{\mathbf{R}}}

\newcommand{\bp}{\mathbf{p}}
\newcommand{\dbp}{\dot{\mathbf{p}}}
\newcommand{\bP}{\mathbf{P}}
\newcommand{\dbP}{\dot{\mathbf{P}}}

\newcommand{\bq}{\mathbf{q}}
\newcommand{\bQ}{\mathbf{Q}}
\newcommand{\dbq}{\dot{\mathbf{q}}}
\newcommand{\dbQ}{\dot{\mathbf{Q}}}

\newcommand{\bPi}{\boldsymbol{\Pi}}

\newcommand{\bA}{\mathbf{A}}
\newcommand{\dbA}{\dot{\mathbf{A}}}
\newcommand{\bcA}{\boldsymbol{\mathcal{A}}}

\newcommand{\bcP}{\boldsymbol{\mathcal{P}}}

\newcommand{\bE}{\mathbf{E}}
\newcommand{\bB}{\mathbf{B}}
\newcommand{\bj}{\mathbf{j}}
\newcommand{\bd}{\mathbf{d}}

\newcommand{\Hc}{\mathrm{H.\,c.}}

\newcommand{\id}{\mathbb{1}}

\begin{document}

\title{Mass-energy and anomalous friction in quantum optics}

\author{Matthias Sonnleitner}
\affiliation{Institute for Theoretical Physics, University of Innsbruck, Technikerstr. 21a, 6020 Innsbruck, Austria}

\author{Stephen M. Barnett}
\affiliation{School of Physics and Astronomy, University of Glasgow, Glasgow G12~8QQ, United Kingdom}

\begin{abstract}
	The usual multipolar Hamiltonian for atom-light interaction features a non-relativistic moving atom interacting with electromagnetic fields which inherently follow Lorentzian symmetry. This combination can lead to situations where atoms appear to experience a friction force, when in fact they only change their internal mass-energy due to the emission or absorption of a photon. Unfortunately, the simple Galilean description of the atom's motion is not sufficient to distinguish between a change in momentum due to acceleration and a change in momentum due to a change in internal mass-energy. In this work we show how a low-order relativistic correction can be included in the multipolar atom-light Hamiltonian. We also give examples how this affects the most basic mechanical interactions between atoms and photons.
\end{abstract}

\maketitle

\section{Introduction}

The mechanical interaction between atoms and light is well studied and successfully tested to high precision, yet there is always room for puzzles and surprises. For example, we recently demonstrated that an excited atom travelling through empty space appears to experience a friction force connected to its spontaneous decay~\cite{sonnleitner2017will,barnett2018vacuum,barnett2018vacuumContPhys}. Of course, a decaying atom will experience a recoil, but in the atom's rest frame one clearly sees that this recoil averages to zero. However, a combination of the first-order Doppler shift and aberration leads to an average change in the atom's momentum $\dot{\ew{\mathbf{P}}} = - \Gamma \hbar k_A \ew{\mathbf{P}}/(M c)$, where $\Gamma$ is the atom's decay rate, $k_A=\omega_A/c$ the wavenumber of the atomic transition, $\ew{\mathbf{P}}$ the momentum expectation value and $M$ the mass of the particle. This change in momentum is thus proportional to the recoil of the particle and its velocity and has the form of a friction force. At first sight this result suggests that one observes a decelerating atom in the laboratory frame, while an observer co-moving with the atom sees no change - this result would contradict both Galilean and Einsteinian relativity. We should point out, for the sake of clarity, that the effect considered here is distinct from the (quantum) frictional force experienced by an atom or other objects moving with respect to a dielectric plate~\cite{polevoi1990tangential,pieplow2015cherenkov,pendry1997shearing,pendry2010quantum,buhmann2013dispersionII}.

The famous connection between energy and inertia, $E=mc^2$~\cite{einstein1905tragheit} solves the puzzle: During the emission process the atom loses energy and hence the momentum changes, not due to a deceleration, but due to a change of mass-energy. It might seem surprising that Einstein's famous formula is required to discuss an effect appearing in \emph{non-relativistic} atom-light interaction. But one can show that the first-order Doppler shift and aberration,\footnote{The radiation emitted by a moving atom is not only Doppler shifted, its emission profile is also changed due to aberration (this can be also understood as lowest order ``relativistic beaming''). Both effects are required to correctly describe the loss of energy for moving radiating objects~\cite{barnett2018vacuum}.} which are both correctly described in the electric dipole Hamiltonian, are enough to obtain these spurious friction terms due to the change in mass-energy~\cite{sonnleitner2017will,barnett2018vacuum}.

Both the puzzle of this ``vacuum friction'' and its solution are classical problems and appear trivial when discussed in a relativistic framework involving four-vectors etc. Here, however, we are interested in the implications of this puzzle in the context of the non-relativistic discussion usually applied in quantum optics. In this context the solution to the friction puzzle had to be included ``by hand''. The framework of the Hamiltonian gave no reason to believe that the mass $M$ appearing in $\mathbf{P} \approx M \dot{\mathbf{R}}$ should be understood as a time-dependent mass-energy. Indeed, the Heisenberg equations of motion for the velocity of the atom does not include this correction. The solution presented in~\cite{sonnleitner2017will} is plausible and strongly supported by classical arguments~\cite{einstein1905tragheit}, but it has not been derived. Other authors noting related effects in moving dielectrics also had to argue similarly~\cite{polevoi1990tangential,pieplow2015cherenkov}.

For consistency the Hamiltonian structure of non-relativistic quantum mechanics should indicate clearly if a change in momentum $\dot{\mathbf{P}}$ is also connected to an acceleration $\ddot{\mathbf{R}}$. After all, the canonical momentum~$\bP$ and the position~$\bR$ are independent variables. In the example of the emitting atom~\cite{sonnleitner2017will}, however, we found $M \ew{\ddot{\mathbf{R}}} = \ew{\dot{\mathbf{P}}}$, wrongly suggesting that the spurious momentum change leads to a friction-like deceleration of the atom.

It should be possible to distinguish, in principle, between actual (friction) forces in atom-light interaction, and those changes in momentum that originate in a change of internal energy and do not lead to a change in the motion of the particle. Given the ever growing precision demanded from and delivered by quantum optics experiments, there may come a time when such a distinction is necessary to explain observations.

In this work we derive a modification of the usual multipolar atom-light Hamiltonian to address these issues. Usually, atom-light interactions are derived from a Lagrangian in which the particles' motion is fully described in a non-relativistic framework. Our approach starts with the Darwin Lagrangian (section~\ref{sec:_DarwinLagrangian}), which gives an approximately relativistic description of the motion of and interaction between charged particles. Canonical quantisation, in section~\ref{sec:_QHamiltonian}, followed by a Power-Zienau-Woolley (PZW) transformation~\cite{power1959coulomb,woolley1971molecular,babiker1983derivation} in section~\ref{sec:_PZW-Trafo} will then give a multipolar description in the center of mass frame. As the usual center of mass is not suitable for an approximately relativistic framework we will perform a further transformation to a central coordinate similar to the center of energy in section~\ref{sec:_final_can_trafo}. There we will arrive at a simple and natural Hamiltonian~\eqref{eq:_Hamiltonian_final}, which is still essentially non-relativistic, but properly accounts for the leading-order mass defect.

The individual steps used in this work have been published before, either in textbooks~\cite{jackson1975classical,landaulifshitzclassical,cohen1989photons} or as individual articles (we are mostly indebted to the work in~\cite{close1970relativistic,lembessis1993theory}) but, to our knowledge, there has not yet been a comprehensive work combining all these aspects and focussing on the role of internal energy in an atom's equations of motion.

Of course, relativistic extensions and corrections have long been part of atomic and molecular physics~\cite{dirac1928quantum,sobelman1992atomic,bethe2008quantum}. To our knowledge, however, these corrections usually consider a static nucleus providing a central potential for the electrons, with the atom as a whole at rest. One reason for this is that the Currie-Jordan-Sudarshan No-Go Theorem states that a fully relativistic canonical description of a set of interacting point particles is impossible, unless one works in a frame with vanishing total momentum~$\bP=0$~\cite{currie1963relativistic,peres2002quantum}. For the problem discussed here it is necessary to give a framework where the total atom is moving. This is possible because our discussion is not fully relativistic but only a lowest-order post-Newtonian extension for which it is possible to construct a Hamiltonian formalism. 

During the completion of the present work we noticed a series of publications by David Alba and co-workers who exhaustively address the issue of relativistic atomic physics, see refs.~\cite{alba2010towards,alba2009towards,alba2013relativistic} and references therein. The low-energy limits of their results also contain modifications to the total kinetic energy of the atom similar to those derived here, \cf section~\ref{sec:_expected_Hamiltonian} or equ.~\eqref{eq:_Hamiltonian_final}. The quantum-optics-inspired approach presented here complements earlier and more rigorous mathematical treatments.

Recently, there has been new interest in studying the coupling between gravity and quantum mechanics in light of Einstein's equivalence principle~\cite{unruh1979bohr,zych2011quantum,zych2018quantum,pikovski2015universal,krause2017relativistic}. Using arguments such as the one given in section~\ref{sec:_expected_Hamiltonian} or based on low-energy limits of the Klein-Gordon equation~\cite{lammerzahl1995hamilton,pikovski2015universal}, one can show that the gravitational field couples not only to the bare mass of an atom, but also to its internal energy state. This coupling is comparable to the corrected Hamiltonian derived in this work, although anomalous friction due to changes in internal energy has not been studied previously.

At the end of this work we shall explore various examples to study the implications of this modification in section~\ref{sec:_Examples}.

\subsection{Expected form of an approximately relativistic Hamiltonian}\label{sec:_expected_Hamiltonian}

Even without a formal derivation it is not difficult to anticipate the type of Hamiltonian we should find later. The classical relativistic Hamiltonian for a free spinless particle of mass $M$ and momentum $\bP$ is given by 
\begin{equation}
	H_\text{free}= \sqrt{ \bP^2 c^2 + M^2 c^4} \approx M c^2 + \frac{\bP^2}{2 M}
	\,.
\end{equation}
We are searching for a Hamiltonian where inertia depends on the internal energy of the particle. Hence, we replace $M\rightarrow M + E_\text{int}/c^2$ such that
\begin{equation}
	H_\text{free} \rightarrow H \approx M c^2 + E_\text{int} + \frac{\bP^2}{2 M}\left(1-\frac{E_\text{int}}{Mc^2}\right)
	\,.
\end{equation}
If $\bQ$ denotes the position of the particle, then this gives $\dbQ = \tpder{\bP}{} H$  or $\bP \approx \dbQ \left(M + E_\text{int}/c^2\right)$. A change of momentum can thus be due to an acceleration or a change in internal energy.

For an atom interacting with an external light field this suggests a Hamiltonian of a form
\begin{equation}\label{eq:_Hamiltonian_expected}
	\mathrm{H} = \frac{\bP^2}{2 M}\left(1 - \frac{\mathrm{H}_\text{A}}{M c^2}\right) + \mathrm{H}_\text{A} + \mathrm{H}_\text{L} + \mathrm{H}_\text{AL}
	\,,
\end{equation}
where $\mathrm{H}_\text{A}$ describes the internal atomic degrees of freedom, $\mathrm{H}_\text{L}$ is the energy of the radiation field and $\mathrm{H}_\text{AL}$ gives the atom light-interaction and the constant mass term has been dropped. In the following sections we shall derive this Hamiltonian.

As mentioned in the introduction, several works on the coupling between internal energy states and gravitational fields use similar arguments~\cite{unruh1979bohr,zych2011quantum,zych2018quantum,pikovski2015universal,krause2017relativistic,yudin2018mass}. But this short derivation considers the atom as a single entity and thus cannot incorporate the subtleties that arise in the microscopic description following below.

\subsection{Hierarchy of energy scales}\label{sec:_energy_hierarchy}
Before we begin to derive our Hamiltonian, let us briefly discuss the energy scales involved in this problem. The biggest energies by far are the total mass and relative mass of the system, $M=m_1+m_2$ and $\mu=m_1 m_2/M$, respectively. Next are the internal transition energies of the atom which are much larger than the energies involved in the atom-light coupling. Table~\ref{tab:_energyscales} gives an overview of these energies for a hydrogen-like setup (particle~1 is an electron, particle~2 the proton).

\begingroup
\begin{table}[h]
\centering
\begin{tabular}{l l}
\hline
	Masses: & $M c^2 \sim \mathrm{GeV}$, $\mu c^2 \sim 0.5\,\mathrm{MeV}$
	\\
	Atomic energies: & $\mathrm{H}_\text{A} \approx \frac{\bp^2}{2\mu} - \frac{e^2}{4\pi\varepsilon_0 r} \sim \mathrm{eV}$
	\\
	Atom-light interaction: \quad & $\mathrm{H}_\text{AL} \approx -\bd\cdot\bE = \hbar \Omega \sim 10^{-9}\mathrm{eV}$
	\\
\hline
\end{tabular}
\caption{\label{tab:_energyscales}%
Comparison of energy scales for a (hydrogen) atom interacting with a radiation field ($M\sim$ proton mass, reduced mass $\mu\sim$ electron mass). The strength of the atom-light interaction is chosen such that the interaction energy of the laser field~$\hbar \Omega$ is of the same order as a typical atomic decay rate $\Omega\sim\Gamma\sim \mathrm{MHz}$.}
\end{table}
\endgroup

Table~\ref{tab:_energyscales} does not include the total kinetic energy of the atom, $\bP^2/(2M)$, which can vary between $10^{-14}\,\mathrm{eV}$ for an ultra cold gas and almost be of the order of atomic energies for a hot vapor. For the approximations made in this work, we shall keep corrections of order $\mathrm{H}_\text{A}/(m_i c^2)$, but drop terms $\sim \mathrm{H}_\text{AL}/(m_i c^2)$, for any of the two masses $i=1,2$.

We note that neglecting terms of order~$\mathrm{H}_\text{AL}/(m_i c^2)$ also implies that we drop any terms possibly giving an Aharonov-Casher interaction, which would couple the magnetic dipole moment and velocity of the atom with an external electric field~\cite{aharonov1984topological}. As shown in equ.~\eqref{eq:_magnetisation}, the magnetic dipole moment already depends on particle velocities, such that the combination of these three terms goes beyond our level of approximation. Studying the relation between our approach and the proposed connection between the Aharonov-Casher effect and the difference between center of mass and center of energy~\cite{horsley2006power} will thus be left for future work.

\section{Classical, approximately relativistic Lagrangian for charges and external radiation fields}\label{sec:_DarwinLagrangian}

We start from the classical Lagrangian for two particles of mass $m_i$, charge $e_i$ at positions $\br_i$ and velocities $\dbr_i$, $i=1,2$, interacting with electromagnetic potentials $\phi_\text{tot}$, $\bA_\text{tot}$~\cite{jackson1975classical},
\begin{equation}\label{eq:_L_initial}
	L = - \sum_{i=1,2} m_i c^2 \sqrt{1-\dbr_i^2/c^2} 
		+ \frac{\varepsilon_0}{2} \int d^3\bx \big( (\partial_t \bA_\text{tot} + \nabla \phi_\text{tot})^2 - c^2 (\nabla\times\bA_\text{tot})^2 \big)
		+ \int d^3\bx \big( \mathbf{j}\cdot \bA_\text{tot} - \rho \phi_\text{tot})
		\,.
\end{equation}
The first term describes the relativistic motion of the particles, although only terms up to~$\sim \dbr_i^2/c^2$ are required to obtain the Hamiltonian outlined above. The second and third terms contain the dynamics of the fields and their interaction with the charges via the density $\rho(\bx,t) = \sum_i e_i \delta(\bx-\br_i(t))$ and current $\mathbf{j}(\bx,t) = \sum_i e_i \dbr_i \delta(\bx-\br_i(t))$.

The potentials $\bA_\text{tot}$ and $\phi_\text{tot}$ contain contributions from external fields as well as the fields generated by the moving charges. With $\bA_\text{tot} = \bcA + \bA$ we shall only keep the external radiation potential $\bA$ as an independent variable, the potential generated by the charges, $\bcA$, shall be expressed in terms of the particle coordinates, \cf appendix~\ref{appendix:_internal fields}. Additionally we choose the Coulomb gauge where $\nabla\cdot\bA = 0 \leftrightarrow \bA \equiv \bA^\perp$ while $\phi_\text{tot}=\phi_\text{internal} \equiv \phi$ (the external scalar potential vanishes in the absence of external charges).\footnote{Recall that any square integrable vector field can be decomposed into longitudinal and transverse components $\mathbf{a}(\bx) = \mathbf{a}^\parallel(\bx) + \mathbf{a}^\perp(\bx)$ with $\nabla\times\mathbf{a}^\parallel=0$ and $\nabla\cdot\mathbf{a}^\perp=0$. Also, recall that $\int d\bx\, \mathbf{a}^\perp \cdot \mathbf{b}^\parallel =0$ such that $\int d\bx\, \bA^\perp \cdot\bj = \int d\bx\, \bA^\perp \cdot\bj^\perp$ and $\int d\bx\, \bA^\perp \cdot \nabla \phi = 0$~\cite{power1964introductory,milonni1994quantum}.} 

The internal potentials satisfy Maxwell's equations in the Coulomb gauge, 
\begin{equation}
	\nabla^2\phi = -\rho/\epsilon_0
	\,,
	\qquad
	\big(\nabla^2 - \tfrac{1}{c^2} \tpder{t}{2}\big) \bcA^\perp = -\mu_0 \bj^\perp
	\,,
\end{equation}
with $\bj^\perp = \bj - \bj^\parallel = \bj - \varepsilon_0 \nabla \tpder{t}{}\phi$. Using these, partial integration and the identity $\nabla\times(\nabla\times \bcA^\perp) = - \nabla^2 \bcA^\perp = \mu_0 \bj^\perp - \partial_t^2 \bcA^\perp/c^2$ we get
\begin{subequations}\label{eq:_L_initial_to_L_intermediate}
\begin{align}
	\frac{\varepsilon_0}{2} \int d^3\bx (\partial_t \bcA^\perp + \nabla \phi)^2 
		&= \frac{\varepsilon_0}{2} \int d^3\bx \big( (\partial_t \bcA^\perp)^2 - (\phi \nabla^2\phi)\big) 
			= \frac{1}{2} \int d^3\bx \big( \varepsilon_0 (\partial_t \bcA^\perp)^2 + \rho \phi \big) 
	\,,	\\
	- \frac{\varepsilon_0 c^2 }{2} \int d^3\bx (\nabla\times \bcA^\perp)^2 
		&= - \frac{1}{2\mu_0} \int d^3\bx\, \bcA^\perp \cdot \big(\nabla\times(\nabla\times \bcA^\perp)\big)
		\notag \\
		&= -\frac{1}{2} \int d^3\bx\, \bcA^\perp \cdot \bj^\perp+ 
		\frac{\varepsilon_0}{2} \int d^3\bx\, \bcA^\perp \cdot (\partial_t^2 \bcA^\perp)
	\,.	
\end{align}
\end{subequations}

Using these relations and $\bcA^\perp \cdot \partial_t^2 \bcA^\perp + (\partial_t \bcA^\perp)^2 = \partial_t^2(\bcA^\perp)^2/2$ we can rewrite the initial Lagrangian from~\eqref{eq:_L_initial} in the form
\begin{multline}\label{eq:_L_intermediate}
	L = \sum_{i=1,2} \frac{m_i \dbr_i^2}{2}\left(1+\frac{\dbr_i^2}{4c^2}\right)
		+ \frac{1}{2}\int d^3\bx \big( \mathbf{j}\cdot \bcA^\perp - \rho \phi)
		+ \frac{\varepsilon_0}{2} \int d^3\bx \big( (\partial_t \bA^\perp)^2 - c^2 (\nabla\times\bA^\perp)^2 \big)
		+ \int d^3\bx\, \mathbf{j}\cdot \bA^\perp
		\\
		+\frac{\varepsilon_0}{4} \int d^3\bx \tpder{t}{2} (\bcA^\perp)^2
		+ \varepsilon_0 \int d^3\bx \big((\partial_t \bcA^\perp)\cdot(\partial_t \bA^\perp) - c^2 (\nabla \times \bcA^\perp)\cdot(\nabla\times\bA^\perp) \big)
\end{multline}
The first two terms of this Lagrangian describe the approximately relativistic motion of the particles and their interaction with the fields generated by their respective counterparts. Using the internal fields calculated in appendix~\ref{appendix:_internal fields} we see that these two terms give the well known Darwin Lagrangian~\eqref{eq:_Lagrangian_Darwin}~\cite{darwin1920dynamical,jackson1975classical}. The third and fourth terms describe the external radiation field and its interaction with the moving charges in the usual form. The remaining terms in the second line are studied in appendix~\ref{appendix:_evil_terms} where we argue why these can be neglected for the problem of interest here.

\section{From classical Lagrangian to quantum Hamiltonian in minimal coupling form}\label{sec:_QHamiltonian}
In the previous section we derived the Lagrangian for two charges with spin zero (the atom) interacting with an external radiation field to the lowest approximately relativistic order,
\begin{equation}\label{eq:_Lagrangian_final}
	L(\br_1,\dbr_1, \br_2,\dbr_2, \bA^\perp, \dbA^\perp) = L_\text{Darwin}(\br_1,\dbr_1, \br_2,\dbr_2) 
		+ \frac{\varepsilon_0}{2} \int d^3\bx \big( (\partial_t \bA^\perp)^2 - c^2 (\nabla\times\bA^\perp)^2 \big)
		+ \int d^3\bx\, \mathbf{j}\cdot \bA^\perp
\,.		
\end{equation}
Using the internal potentials~$\phi$ and~$\bcA^\perp$ calculated in appendix~\ref{appendix:_internal fields}, \cf equs.~\eqref{eq:_phi_internal} and~\eqref{eq:_Aperp_internal}, we obtain the Darwin Lagrangian~\cite{darwin1920dynamical}
\begin{multline}\label{eq:_Lagrangian_Darwin}
	L_\text{Darwin}(\br_1,\dbr_1, \br_2,\dbr_2) = 
	\sum_{i=1,2} \frac{m_i \dbr_i^2}{2}\left(1+\frac{\dbr_i^2}{4c^2}\right) + \frac{1}{2}\int d^3\bx \big( \mathbf{j}\cdot \bcA^\perp - \rho \phi)
	\\
	= \frac{m_1 \dbr_1^2}{2} + \frac{m_1 \dbr_1^4}{8 c^2} 
	+ \frac{m_2 \dbr_2^2}{2} + \frac{m_2 \dbr_2^4}{8 c^2}
		- \frac{1}{4 \pi \varepsilon_0}\frac{e_1 e_2}{r} \left(1- \frac{\dbr_1\cdot \dbr_2}{2 c^2}\right)
		+ \frac{e_1 e_2}{4 \pi \varepsilon_0} \frac{ (\dbr_1 \cdot \br) (\dbr_2 \cdot \br)}{2 r^3 c^2}
		\;,
\end{multline}
where $\br = \br_1-\br_2$ and $r=\abs{\br}$. If we define $\br_1$ and $\br_2$ to be the positions of the electron and proton, respectively, then $e_2=-e_1 = e$, the elementary charge, and $\br$ is the vector pointing from the nucleus to the electron.

To obtain the corresponding (classical) Hamiltonian we calculate the canonical momenta associated with the particle coordinates $\br_i$ and the external field~$\bA^\perp$,
\begin{align}
	\bp_i &= \pder{\dbr_i}{} L 
		= m_i \dbr_i + e_i \bA^\perp(\br_i) - m_i \br_i \frac{\dbr_i^2}{2c^2} + \frac{1}{4\pi\varepsilon_0} \frac{e_1 e_2}{2 r c^2} \left( \dbr_j +  \frac{\br (\dbr_j\cdot \br)}{r^2}\right) 
	\,, \\
	\bPi^\perp &= \pder{\text{$\dbA^\perp$}}{} L 
		= \varepsilon_0 \dbA^\perp
	\,,	
\end{align}
for $i=1,2$ and $j\neq i$. This gives us the resulting classical Hamiltonian
\begin{multline}
	H= \frac{\overline{\bp}_1^2}{2 m_1} + \frac{\overline{\bp}_1^4}{8 m_1^3 c^2} 
		+ \frac{\overline{\bp}_2^2}{2 m_2} + \frac{\overline{\bp}_2^4}{8 m_2^3 c^2} 
		+ \frac{1}{4 \pi \varepsilon_0}\frac{e_1 e_2}{r} \left(1- \frac{\overline{\bp}_1\cdot \overline{\bp}_2}{2 m_1 m_2 c^2}\right)
		- \frac{e_1 e_2}{4 \pi \varepsilon_0} \frac{ (\overline{\bp}_1 \cdot \br) (\overline{\bp}_2 \cdot \br)}{2 r^3 c^2}
		\\
		+ \frac{\varepsilon_0}{2} \int d^3\bx \big( (\bPi^\perp/\varepsilon_0)^2 + c^2 (\nabla\times\bA^\perp)^2 \big)
	\,,	
\end{multline}
where $\overline{\bp}_i := \bp_i + e_i \bA^\perp(\br_i)$.

We see that this classical Hamiltonian contains products of the momenta $\bp_i$ with functions of the positions $\br_i$. As the corresponding quantum operators do not commute we have to be careful to obtain the correct ordering during the canonical quantisation. To achieve this it is useful to return to the Darwin Lagrangian~\eqref{eq:_Lagrangian_Darwin} to see that these problematic terms arise from the coupling $\bj\cdot\bcA^\perp$, which can be symmetrized as
\begin{align*}
	\frac{1}{2} \int d^3\bx\, \bj\cdot\bcA^\perp
		& = \frac{1}{4} \left[ e_1 \left(\dbr_1\cdot\bcA^\perp(\br_1) + \bcA^\perp(\br_1)\cdot \dbr_1 \right)
			+  e_2 \left(\dbr_2\cdot\bcA^\perp(\br_2) + \bcA^\perp(\br_2)\cdot\dbr_2\right) \right]
	\\
		&\approx
		\frac{1}{4} \left[ \frac{e_1}{m_1} \left(\overline{\bp}_1\cdot\bcA^\perp(\br_1) +  \bcA^\perp(\br_1)\cdot\overline{\bp}_1\right) + (1\leftrightarrow2) \right]
	\\
		&= 
		\frac{e_1 e_2}{16 \pi \varepsilon_0 c^2 m_1 m_2}
			\left[ \overline{\bp}_1 \cdot \frac{1}{r} \overline{\bp}_2 + (\overline{\bp}_1 \cdot \br) \frac{1}{r^3} (\br\cdot\overline{\bp}_2) + (1\leftrightarrow2) \right]
\,.			
\end{align*}
For the final line we used the fact that $[\bp_1, \br]=-[\bp_2,\br]$ and a symmetric version of $\bcA^\perp$ from equ.~\eqref{eq:_Aperp_internal},
\begin{equation*}
	\bcA^\perp(\br_i) = \frac{e_j}{16 \pi \varepsilon_0 c^2 m_j} \left[ \frac{1}{r}\overline{\bp}_j + \overline{\bp}_j\frac{1}{r} + \frac{\br}{r^3} (\br \cdot \overline{\bp}_j) + (\overline{\bp}_j \cdot \br)\frac{\br}{r^3}  \right]
	\,,
\end{equation*}
with $j\neq i$ to avoid infinite interaction between particles and their own fields.

The quantum Hamiltonian in minimal coupling form thus becomes
\begin{multline}\label{eq:_H_indivparticles}
	\mathrm{H}_\text{[Min.c.]}= \frac{\overline{\bp}_1^2}{2 m_1} + \frac{\overline{\bp}_2^2}{2 m_2}
		+ \frac{1}{4 \pi \varepsilon_0}\frac{e_1 e_2}{r}
		+ \frac{\varepsilon_0}{2} \int d^3\bx \big[ (\bPi^\perp/\varepsilon_0)^2 + c^2 (\nabla\times\bA^\perp)^2 \big]
		\\
		+ \frac{\overline{\bp}_1^4}{8 m_1^3 c^2}  + \frac{\overline{\bp}_2^4}{8 m_2^3 c^2} 
		- \frac{e_1 e_2}{16 \pi \varepsilon_0 c^2 m_1 m_2}
			\left[ \overline{\bp}_1 \cdot \frac{1}{r} \overline{\bp}_2 + (\overline{\bp}_1 \cdot \br) \frac{1}{r^3} (\br\cdot\overline{\bp}_2) + (1\leftrightarrow2) \right]
			\,,
\end{multline}
where the first line is the usual expression for two charged particles in a radiation field and the second line is due to post-Newtonian relativistic corrections. A similar setup including the spin of the electron but without external fields can be derived from the Dirac equation and is then called the Breit-Hamiltonian~\cite{breit1932dirac,bethe2008quantum}. Of course, including the spin of the electron and nucleus gives rise to a new wealth of phenomena including, for example, \emph{Zitterbewegung} and associated subtleties in the interpretation of position and momentum operators~\cite{foldy1950dirac,cameron2018relativistic}.

Note that from now on $\br_i$, $\bp_i$, $\bA^\perp$ and $\bPi^\perp$ are understood to be operators with $[\br_{i,k}, \bp_{j,l}] = i \hbar \delta_{i,j}\delta_{k,l}$ for $\br_{i,k}$ being the $k$th component for the position operator of the $i$th particle and $[\bA_k^\perp(\bx),\bPi_l^\perp(\bx')] = i \hbar \delta_{k,l} \delta^\perp(\bx-\bx')$~\cite{loudon2000quantum}. But aside from the examples in section~\ref{sec:_Examples}, the entire derivation given here is also valid for classical systems where the commutators are replaced by Poisson brackets.

\section{PZW transformation to a multipolar Hamiltonian in center of mass coordinates}\label{sec:_PZW-Trafo}

The Hamiltonian given in equation~\eqref{eq:_H_indivparticles} describes the approximately relativistic dynamics of two charged particles of spin zero in the presence of external radiation fields. As we are interested in a separation between relative and central dynamics of the particles, we shall transform the minimal coupling Hamiltonian $\mathrm{H_\text{[Min.c.]}}$ to a multipolar form~\cite{power1964introductory}.

There are a number of ways to transform a minimal coupling Hamiltonian into a form with central and relative coordinates suitable to describe mobile atoms or molecules~\cite{cohen1989photons}. Here we choose the Power-Zienau-Woolley (PZW) transformation in the form of a unitary transformation of the Hamiltonian combined with a multipolar expansion and the introduction of center-of-mass coordinates. But of course it would also be possible to apply an equivalent transformation to the underlying Lagrangian~\cite{babiker1983derivation,boussiakou2002electrodynamics,konya2018equivalence}, similar to the Göppert-Mayer transformation~\cite{goppert1931elementarakte}.

The discussion here follows closely that given in previous works~\cite{lembessis1993theory,loudon2000quantum} and is included here for completeness. The PZW transformation is given by the unitary operator,
\begin{equation}\label{eq:_PZW_unitary}
	U = e^{- i \Lambda} = \exp\left[ \frac{i}{\hbar} \int d^3\bx \, \bcP(\bx,t)\cdot \bA^\perp(\bx,t) \right]
	\,,
\end{equation}
where $\bcP$ is the polarisation field centered at a position~$\bR$,
\begin{equation}\label{eq:_polarisation}
	\bcP(\bx,t) = \sum_{i=1,2} e_i (\br_i(t) - \bR(t)) \int_0^1 d\lambda \,\delta( \bx-\bR(t) - \lambda(\br_i(t)-\bR(t)))
	\,.
\end{equation}

The interpretation of $\bcP$ as the polarisation field is supported by $\nabla\cdot \bcP(\bx,t) = -\rho (\bx, t)$, where $\rho$ is the density of the (bound) charges in the atom such that the electric displacement field is $\mathbf{D} = \varepsilon_0 \bE + \bcP = -\bPi + \bcP$~\cite{konya2018equivalence}. We also get 
\begin{equation}\label{eq:_polfield_timeder}
	\partial_t \bcP(\bx,t) = \bj(\bx,t) - \nabla\times\boldsymbol{\mathcal{M}}(\bx,t) + \nabla\times\left( \dbR \times \bcP(\bx,t) \right)
\end{equation}
with the magnetisation field
\begin{equation}\label{eq:_magnetisation}
	\boldsymbol{\mathcal{M}}(\bx,t) = \sum_{i=1,2} e_i (\br_i(t)-\bR(t)) \times (\dbr_i(t)-\dbR(t)) \int_0^1 d\lambda \,\lambda \delta( \bx-\bR(t) - \lambda(\br_i(t)-\bR(t)))\,.
\end{equation}
The final term in equ.~\eqref{eq:_polfield_timeder} is the so-called Röntgen current which describes the magnetic moment due to an electric polarisation moving at velocity $\dbR$~\cite{rontgen1888ueber,hnizdo2012magnetic,sonnleitner2017rontgen}. In principle, $\bR$ could be any position we choose to be the center of our atom, but as discussed in section~\ref{sec:_Q_in_PZW} it is convenient to choose the center of mass,
\begin{equation}\label{eq:_com}
	\bR = \frac{m_1 \br_1 + m_2 \br_2}{M}\,,
\end{equation}
where $M=m_1+m_2$.

As $\Lambda$ is a function of the positions $\br_i$ and the potential $\bA^\perp$, only the respective canonical momenta are changed:
\begin{subequations}
\begin{align}
	\bp_i &\rightarrow U \bp_i U^\dagger = \bp_i + \hbar \nabla_{\br_i} \Lambda
	\,, \\
	\bPi^\perp(\bx) &\rightarrow \bPi^\perp(\bx) - \bcP^\perp(\bx)
\,.	
\end{align}
\end{subequations}
Here, $\nabla_{\br_i}$ denotes derivation with respect to the position of the $i$th particle.

Performing the integral with respect to $d^3\bx$ shows that $\hbar \Lambda = - \sum_j e_j \int_0^1 d\lambda \,\overline{\br}_j \cdot \bA^\perp(\bR + \lambda\overline{\br}_j)$, where $\overline{\br}_j:= \br_j-\bR$. Expanding to first order in $\overline{\br}_j$ (electric dipole approximation\footnote{There is no particular reason to stop the multipole-expansion at the level of dipoles. Expanding to higher orders of~$\overline{\br}_j$ would lead to the usual quadrupole etc. terms with corresponding corrections due to the center of mass motion~\cite{babiker2002orbital}.}) we obtain
\begin{align}
	\nabla_{\br_i} \overline{\br}_j \cdot \bA^\perp(\bR + \lambda\overline{\br}_j) 
	&= \left(\delta_{ij} - \tfrac{m_i}{M}\right) \bA^\perp(\bR + \lambda\overline{\br}_j) 
	\notag \\ &\qquad
	+ \left(\lambda \delta_{ij} + (1-\lambda) \tfrac{m_i}{M}\right) \left[ \big(\overline{\br}_j\cdot\nabla\big) \bA^\perp(\bR + \lambda\overline{\br}_j) + \overline{\br}_j\times(\nabla \times \bA^\perp(\bR + \lambda\overline{\br}_j)) \right]
	\notag \\
	&\simeq
	\left(\delta_{ij} - \tfrac{m_i}{M}\right) \left[\bA^\perp(\bR) + \lambda \big(\overline{\br}_j\cdot\nabla\big) \bA^\perp(\bR) \right]
	\notag &\\ &\qquad
	+ \left(\lambda \delta_{ij} + (1-\lambda) \tfrac{m_i}{M}\right) \left[ \big(\overline{\br}_j\cdot\nabla\big) \bA^\perp(\bR) + \overline{\br}_j\times(\nabla \times \bA^\perp(\bR)) \right]
	\,.
\end{align}
Integrating over $\lambda$ and using $\sum_{j=1,2} e_j = 0$, we find
\begin{equation}
	\hbar \nabla_{\br_{1,2}} \Lambda \simeq \pm e \left(\bA^\perp(\bR) + (\overline{\br}_{1,2}\cdot\nabla)\bA^\perp(\bR)\right) + \frac{e_1 \br_1 + e_2 \br_2}{2} \times (\nabla\times \bA^\perp(\bR)) \,.
\end{equation}
Recognising that $\bA^\perp(\bR) + (\overline{\br_i}\cdot\nabla)\bA^\perp(\bR) \simeq \bA^\perp(\br_i)$ and defining the electric dipole moment $\mathbf{d}= \sum_i e_i \br_i$ we see that the PZW-transformation followed by the dipole approximation gives $\bp_i + e_i \bA(\br_i) \rightarrow \bp_i + \mathbf{d} \times \bB(\bR)/2$. 

Inserting this into the second line of the Hamiltonian from~\eqref{eq:_H_indivparticles} we notice that there will be terms of type
\begin{equation}\label{eq:_neglect_small_AL-interactions}
	\frac{\bp_i \cdot (\mathbf{d} \times \bB(\bR))}{m_i m_j c^2} \propto \frac{\abs{\bp_i}}{m_i c} \frac{\abs{\mathbf{d}\cdot\mathbf{E}(\mathbf{\bR})}}{m_j c^2}
	\,.
\end{equation}
As $\abs{\mathbf{d}\cdot\mathbf{E}(\mathbf{\bR})} \ll e^2/(4\pi\varepsilon_0 r) \ll m_i c^2$ is the energy scale for the interaction between the atom and the field, we can safely neglect these terms so that
\begin{multline}\label{eq:_H_com_intermediate}
	\mathrm{H}_\text{[mult]}\simeq \frac{\left(\bp_1 + \tfrac{1}{2}\mathbf{d} \times \bB(\bR) \right)^2}{2 m_1} + \frac{\left(\bp_2 +  \tfrac{1}{2}\mathbf{d} \times \bB(\bR) \right)^2}{2 m_2}
		- \frac{e^2}{4 \pi \varepsilon_0 r}
		+ \frac{\varepsilon_0}{2} \int d^3\bx \big( (\bPi^\perp - \bcP_d^\perp)^2/\varepsilon_0^2 + c^2 \mathbf{B}^2 \big)
			\\
		+ \frac{\bp_1^4}{8 m_1^3 c^2}  + \frac{\bp_2^4}{8 m_2^3 c^2} 
		+ \frac{e^2}{16 \pi \varepsilon_0 c^2 m_1 m_2}
			\left[ \bp_1 \cdot \frac{1}{r} \bp_2 + (\bp_1 \cdot \br) \frac{1}{r^3} (\br\cdot\bp_2) + (1\leftrightarrow2) \right]
\end{multline}
is the multipolar Hamiltonian $\mathrm{H}_\text{[mult]} = U \mathrm{H}_\text{[Min.c.]} U^\dagger$ in electric dipole approximation, $\bcP_d = - \mathbf{d} \delta(\bx-\bR)$. To complete the change to the center of mass frame, we define the momentum conjugate to the distance variable $\br$ as $\bp_\br$ such that 
\begin{equation}
	\bp_{1,2} = \frac{m_{1,2}}{M} \bP \pm \bp_\br \,,
\end{equation}
where $\bP=\bp_1+\bp_2$ is the total momentum which is also conjugate to~$\bR$. Using $[\bP, \bp_r] = [\bP, \br] = 0$ and the relative mass $\mu=m_1 m_2/M$ we obtain a Hamiltonian in the center of mass frame
\begin{subequations}\label{eq:_H_com}
\begin{align}
	\mathrm{H}_\text{[com]} &= \mathrm{H}_\text{C} + \mathrm{H}_\text{A} + \mathrm{H}_\text{AL} + \mathrm{H}_\text{L} + \mathrm{H}_\text{X}	
\,,\\
	\mathrm{H}_\text{C}&= \frac{\bP^2}{2M}\left[ 1 - \frac{\bP^2}{4 M^2 c^2} - \frac{1}{Mc^2}\left(\frac{\bp_\br}{2 \mu} - \frac{e^2}{4\pi \varepsilon_0 r}\right) \right]
	\label{eq:_H_com_HC}
\,,\\
	\mathrm{H}_\text{A}&= \frac{\bp_\br^2}{2\mu} \left( 1 - \frac{m_1^3 + m_2^3}{M^3} \frac{\bp_\br^2}{4 \mu^2 c^2} \right) 
		-\frac{e^2}{4\pi \varepsilon_0} \left[ \frac{1}{r} + \frac{1}{2 \mu M c^2}\left( \bp_\br\cdot \frac{1}{r} \bp_\br + \bp_\br\cdot\br \frac{1}{r^3} \br\cdot\bp_\br \right)\right]
	\label{eq:_H_com_HA}	
\,,\\
	\mathrm{H}_\text{AL} &= - \bd\cdot\bE^\perp(\bR) + \frac{1}{2M} \left[\bP \cdot (\bd \times \bB(\bR)) + \Hc \right] 
	  \notag \\ &\qquad
	-\frac{m_1-m_2}{2 m_1 m_2}\left[ \bp_\br \cdot (\bd \times \bB(\bR)) + \Hc \right] 
	+ \frac{1}{8 \mu} \left(\bd \times\bB(\bR)\right)^2 + \frac{1}{2\varepsilon_0}\int d^3\bx \, {\bcP_d^\perp}^2(\bx,t)
	\label{eq:_H_com_HAL}
\,,\\
	\mathrm{H}_\text{L} &= \frac{\varepsilon_0}{2} \int d^3\bx \big({\bE^\perp}^2 + c^2 \mathbf{B}^2 \big)
	\label{eq:_H_com_HL}
\,,\\
	\mathrm{H}_\text{X} &=
	 - \frac{(\bP\cdot\bp_\br)^2}{2 M^2 \mu c^2} + \frac{e^2}{4\pi\varepsilon_0 r} \frac{(\bP\cdot \br/r)^2}{2 M^2 c^2}
	 \notag \\ &\qquad
	 +\frac{m_1-m_2}{2 \mu M^2 c^2}\left[ (\bP\cdot\bp_\br) \bp_\br^2/\mu - \frac{e^2}{8\pi\varepsilon_0} \left( \frac{1}{r} \bP\cdot\bp_\br + \frac{1}{r^3} (\bP\cdot\br) (\br\cdot\bp_\br) + \Hc \right)\right]
	 \label{eq:_H_com_HX}
\,.	 
\end{align}
\end{subequations}
Here $\mathrm{H}_\text{C}$ describes the central dynamics and $\mathrm{H}_\text{A}$ is the atomic Hamiltonian with relativistic corrections; neglecting the term~$(\bP/(Mc))^2$ we see that $H_\text{C}$ already contains $\mathrm{H}_\text{A}/(Mc^2)$ and thus has the desired form given in equ.~\eqref{eq:_Hamiltonian_expected}. $\mathrm{H}_\text{AL}$ gives the interaction between the atom and the radiation field in the dipole approximation; the first line contains the usual $\bd\cdot\bE$ coupling as well as the Röntgen term, in the second line we find magnetic dipole terms and higher order couplings which are usually neglected~\cite{lembessis1993theory}.\footnote{The Röntgen term too is usually neglected, but here we keep it as it is essential to describe the emission of radiation by moving dipoles and the associated momentum and mass-energy change. The coupling $ - \bd\cdot\bE$ itself would also lead to a momentum change resembling a ``vacuum friction'', but it is only the corrected emission profile obtained using the Röntgen term which allows the interpretation of a change in mass-energy~\cite{sonnleitner2017will,barnett2018vacuum,barnett2018vacuumContPhys,wilkens1994significance}.} $\mathrm{H}_\text{L}$ gives the energy of the radiation field. 

Finally, $\mathrm{H}_\text{X}$ contains terms that couple the internal degrees of freedom $\br$, $\bp_\br$ with the central momentum of the atom, $\bP$. As we discuss in more detail in section~\ref{sec:_difference_com_coe_Q}, the presence of this term shows that the coordinates~$\bR$ and~$\br$ and their respective momenta are not the optimal choice to separate central and relative dynamics in an approximately relativistic setting. We therefore need a further canonical transformation to more suitable coordinates.

\section{Separation of central and relative dynamics}\label{sec:_final_can_trafo}

Given that a system's energy content is part of its inertia it should not be surprising that the center of mass~$\bR$ is not the optimal choice for an approximately relativistic discussion. This is probably why several textbooks dealing with the Darwin Lagrangian contain the exercise to rewrite the corresponding Hamiltonian in the center of energy frame~\cite{jackson1975classical,landaulifshitzclassical} and for $\bP=0$ the reader will find the classical equivalent to $\mathrm{H}_\text{A}$ given in equ.~\eqref{eq:_H_com_HA}. However, these exercises are problematic because the center of energy coordinate  $\bfR$ will also depend on the particle velocities $\dbr_i$. Changing from $\br_1, \br_2$ to central and relative coordinates $\bfR$ and $\br$ is therefore \emph{not} a point transformation which preserves the dynamics of the single-particle Lagrangian. Such a change is also not a canonical transformation as it does not preserve commutation relations. These problems are only covered up if we choose $\bP=0$. We shall discuss this in more detail in section~\ref{sec:_difference_com_coe_Q} and appendix~\ref{appendix:_proof_no_CoE}

It turns out that finding a canonical set of central and relative coordinates for a relativistic, interacting set of point-particles is a surprisingly complicated task. A theorem by Currie, Jordan and Sudarshan even states that it is impossible to find a relativistic canonical formalism for interacting point-particles with non-vanishing total momentum $\bP\neq0$, if certain general conditions should hold as well~\cite{currie1963relativistic,peres2002quantum}. 

Nevertheless, Close and Osborn managed to find expressions for central and relative coordinates which at least allow for an approximately relativistic discussion of interacting particles~\cite{close1970relativistic}.\footnote{Their result was derived for the Breit-Hamiltonian~\cite{breit1932dirac}, which is essentially the Darwin Hamiltonian for particles with spin~1/2, but it can be reduced to the form required here.} Applying their result to our case we obtain the following transformation to the new coordinates $\bQ, \bq$ with their respective canonical momenta $\bP, \bp$,
\begin{subequations}\label{eq:_can_trafo}
\begin{align}
	\bR &= \bQ + \frac{m_1-m_2}{2 M^2 c^2}\left[ \left( \frac{\bp^2}{2 \mu} \bq + \Hc\right) - \frac{e^2}{4\pi \varepsilon_0 q} \bq\right] - \frac{1}{4 M^2 c^2} \left[ (\bq\cdot\bP) \bp + (\bP\cdot\bp)\bq +\Hc\right]
\,,\\	
	\br &= \bq + \frac{m_1-m_2}{2 \mu M^2 c^2}\left[ (\bq\cdot\bP) \bp + \Hc \right] - \frac{\bq\cdot\bP}{2 M^2 c^2} \bP
\,,\\	
	\bp_\br &= \bp + \frac{\bp\cdot\bP}{2 M^2 c^2} \bP - \frac{m_1-m_2}{2 M^2 c^2}\left[ \frac{\bp^2}{\mu}\bP - \frac{e^2}{4\pi\varepsilon_0}\left( \frac{1}{q}\bP -\frac{1}{q^3} (\bP\cdot\bq) \bq\right)\right]
\,.
\end{align}
\end{subequations}
Note that $\bP$ remains unchanged as it already is the correct total momentum and $q=\abs{\bq}$. These new coordinates satisfy the canonical commutation relations: $[\bQ_k, \bP_l] = [\bq_k, \bp_l] = i\hbar \delta_{kl}$ and $[\bQ_k, \bq_l] = [\bQ_k, \bp_l] = [\bP_k, \bp_l] = [\bP_k, \bq_l] = 0$. The external radiation field $\bA^\perp$ at its canonical momentum $\bPi^\perp$ remain unchanged.

As the differences between the old and new coordinates are small, most terms of the Hamiltonian~\eqref{eq:_H_com} transform by a simple replacement $\bp_\br\rightarrow\bp$ and $\br\rightarrow\bq$. The only exceptions are
\begin{subequations}
\begin{align}
	\frac{\bp_\br^2}{2\mu} &\rightarrow 
		\frac{\bp^2}{2\mu} + \frac{(\bp\cdot\bP)^2}{2 \mu M^2 c^2} 
		- \frac{m_1-m_2}{2 \mu M^2 c^2}\left[ \frac{\bp^2}{\mu}\bP\cdot\bp - \frac{e^2}{8\pi\varepsilon_0}\left(\frac{1}{q} \bP\cdot\bp - \frac{1}{q^3} (\bP\cdot\bq) (\bq\cdot\bp) + \Hc\right)\right]
\,,\\
	-\frac{e^2}{4\pi\varepsilon_0 r} &\rightarrow 
	-\frac{e^2}{4\pi\varepsilon_0 q} + \frac{m_1-m_2}{2 \mu M^2 c^2} \frac{e^2}{4\pi\varepsilon_0} \left( \frac{1}{q^3} (\bP\cdot\bq) (\bq\cdot\bp) + \Hc \right) - \frac{e^2}{4\pi\varepsilon_0 q^3} \frac{(\bP\cdot\bq)^2}{2 M^2 c^2} \,.
\end{align}
\end{subequations}
These changes add up such that the transformation to the new coordinates gives
\begin{equation}
	\mathrm{H}_\text{A}(\br, \bp_\br) + \mathrm{H}_\text{X}(\br, \bp_\br, \bR, \bP) \rightarrow \mathrm{H}_\text{A}(\bq, \bp)
\end{equation}
where $\mathrm{H}_\text{A}(\bq, \bp)$ is the atomic Hamiltonian from equ.~\eqref{eq:_H_com_HA} with $\br$ and $\bp_\br$ replaced by $\bq$ and $\bp$, respectively. The total transformed Hamiltonian thus reads
\begin{equation}\label{eq:_Hamiltonian_final}
	\mathrm{H}= \frac{\bP^2}{2M}\left( 1 - \frac{\mathrm{H}_\text{A}(\bq, \bp)}{Mc^2}\right) + \mathrm{H}_\text{A}(\bq, \bp) + \mathrm{H}_\text{AL}(\bQ,\bP,\bA^\perp,\bPi^\perp) + \mathrm{H}_\text{L}(\bA^\perp,\bPi^\perp)
	\,,
\end{equation}
with $\mathrm{H}_\text{AL}(\bQ,\bP,\bA^\perp,\bPi^\perp) = -\bd\cdot\bE^\perp(\bQ) + \frac{1}{2M} \left[\bP \cdot (\bd \times \bB) + \Hc \right]$ in the electric dipole approximation. 

This Hamiltonian and the path to derive it are the main result of this work. It shows that it is possible to construct a Hamiltonian which gives a clear distinction between the change in momentum and the actual acceleration, even if the internal atomic energy changes. It also has the form anticipated in equ.~\eqref{eq:_Hamiltonian_expected}, which is most satisfactory.

\subsection{Remark: The difference between center of mass, center of energy and~\texorpdfstring{$\bQ$}{Q}}\label{sec:_difference_com_coe_Q}
In the course of this work we mentioned three central coordinates: the center of mass~$\bR$~\eqref{eq:_com}, the new central coordinate~$\bQ$~\eqref{eq:_can_trafo} and the center of energy given by~\cite{landaulifshitzclassical,boyer2005illustrations}
\begin{equation}\label{eq:_coe}
	\left(\mathcal{E}_1+\mathcal{E}_2\right) \bfR = \br_1 \mathcal{E}_1 + \br_2 \mathcal{E}_2
\end{equation}
for $\mathcal{E}_i = m_i c^2 + m_i \dbr_i^2/2 - e^2/(8\pi \varepsilon_0 r)$. As previously, we use an approximation to second order to obtain
\begin{align}
	\bfR &\simeq \bR - \frac{m_1-m_2}{4 M^2 c^2} \left[ \bq \left( \frac{\bp^2}{\mu} - \frac{e^2}{4\pi \varepsilon_0 q} \right) + \Hc\right] + \frac{1}{2 M^2 c^2}\left[ \bq (\bP\cdot\bp) + \Hc\right] 
		\notag \\
		&= \bQ + \frac{1}{4 M^2 c^2}\left[ \bq(\bP\cdot\bp) - \bp(\bP\cdot\bq) + \Hc \right]
		\label{eq:_coe_compared_to_Q}
\,.		
\end{align}
From the definitions of $\bR$ and $\bfR$ we see that both are of a form $\alpha_1 \br_1 + \alpha_2 \br_2$ with $\alpha_1+\alpha_2=1$ (\eg $\alpha_i =m_i/M$ for $\bR$). These points therefore lie on a straight line connecting the particle positions $\br_1$ and $\br_2$. $\bQ$, however, has a component pointing in the direction of $\bp$, it is therefore not even restricted to the plane spanned by $\br_1$ and $\br_2$.

If we consider the dynamics of these central coordinates, then we find that, without external fields, 
\begin{equation}
	\dbR = \tfrac{i}{\hbar} [\mathrm{H}_\text{[com]},\bR] = \tfrac{1}{M}\bP(1-\mathrm{H}_\text{A}/(Mc^2)) + \tfrac{i}{\hbar} [\mathrm{H}_\text{X},\bR]
	\,.
\end{equation}
As we see from equ.~\eqref{eq:_H_com} we also get $ [\mathrm{H}_\text{[com]},[\mathrm{H}_\text{X},\bR]] \neq 0$ such that the center of mass accelerates, even if there is no interaction with an external field and $\dbP=0$. Similarly, also $\mathrm{H}_\text{A}$ changes as $[\mathrm{H}_\text{X},\mathrm{H}_\text{A}]\neq 0$, but these changes do not suffice to cancel the acceleration of~$\bR$. This illustrates why the center of mass is not a suitable central coordinate here.

As the Hamiltonian given in equ.~\eqref{eq:_Hamiltonian_final} decouples central and relative dynamics, this cannot happen for~$\bQ$. From equ.~\eqref{eq:_coe_compared_to_Q} we also find that $\dbfR-\dbQ = \tder{t}{} \left[ \bq(\bP\cdot\bp) - \bp(\bP\cdot\bq) + \Hc \right]/(4 M^2 c^2)$. As the relative coordinates follow the Coulomb potential in the atom we get $\dbp \parallel \bq$ and $\dbq \parallel \bp$ such that $\dbfR-\dbQ = \left[ \bq(\dbP\cdot\bp) - \bp(\dbP\cdot\bq) + \Hc \right]/(4 M^2 c^2)$. However $\dbP$ is given by the external fields and as we decided to drop terms~$\propto \mathrm{H}_\text{AL}/(Mc^2)$ we can conclude that $\dbfR \approx \dbQ$, \cf section~\ref{sec:_energy_hierarchy} or equ.~\eqref{eq:_neglect_small_AL-interactions}. Hence, although $\bQ$ is not the center of energy its dynamics are the same such that all associated conservation laws hold to our level of approximation~\cite{boyer2005illustrations}.

In appendix~\ref{appendix:_proof_no_CoE} we show that a set of coordinates that includes the center of energy~$\bfR$ does not satisfy the canonical commutation relations for independent central and relative coordinates. This means that there is no canonical transformation connecting, for example, the individual particle coordinates $\br_i$ with $\bfR$.

\subsection{Remark: Using the center of mass in the PZW transformation}\label{sec:_Q_in_PZW}
Given that the we replaced the center of mass~$\bR$ by the new central coordinate~$\bQ$ one could argue that it would be more reasonable to use~$\bQ$ instead of~$\bR$ in the Power-Zienau-Wolley transformation as well, \cf section~\ref{sec:_PZW-Trafo} equs.~\eqref{eq:_PZW_unitary} and~\eqref{eq:_polarisation}. The phase~$\Lambda(\br_i, \bA^\perp)$ would then become a function of particle positions as well as their momenta~$\bp_i$, such that the PZW-transformation would not only transform momenta, but also positions $\br_i \rightarrow U \br_i U^\dagger$.

More generally, this approach would spoil the connection between the PZW-transformation and similar transformations at the Lagrangian level of type $L \rightarrow L - \tder{t}{}\Lambda$. Such transformations leave the equations of motion unchanged only if $\Lambda$ is a function of coordinates $\br_i$ and $\bA$, not of their derivatives~\cite{goppert1931elementarakte,babiker1983derivation,konya2018equivalence}.

It thus appears to be more instructive to first use a standard PZW-transformation using the center of mass and then make a separate canonical transformation to the coordinates~$\bq$ and $\bQ$.

\subsection{Remark: Expressing the atomic Hamiltonian in terms of energy eigenstates}
The basic atomic Hamiltonian $\mathrm{H}_\text{A}$ given above will usually be replaced by a more complicated version including spin, multiple electrons etc. In the end, one will obtain the energy eigenstates of this atom, $\mathrm{H}_\text{A} = \sum_n E_n \ketbra{n}{n}$, where $E_n<0$ for a bound state. If the setup is such that the atom can be restricted to a two level system, $\mathrm{H}_\text{A} = E_g \ketbra{g}{g} + E_e \ketbra{e}{e}$, with $E_g<E_e<0$, then one can write
\begin{equation}
	\mathrm{H}_\text{A} = E_g \ketbra{g}{g} + E_e \ketbra{e}{e} = \frac{\hbar \omega_A}{2} \sigma_z + \frac{E_e+E_g}{2} \id \,,
\end{equation}
where $E_e-E_g = \hbar \omega_A$ and $\sigma_z = \ketbra{e}{e} - \ketbra{g}{g}$. For an isolated atomic Hamiltonian the constant final term can be dropped. For the modified kinetic-energy term this implies
\begin{equation}
	\frac{\bP^2}{2M}\left( 1 - \frac{\mathrm{H}_\text{A}}{Mc^2}\right) \approx \frac{\bP^2}{2 M'}\left( 1 -\frac{\hbar \omega_A}{2 M' c^2} \sigma_z\right)
	\,.
\end{equation}
Here $M' = [M + (E_e-E_g)/(2 c^2)]$, where $M$ is the mass of the atom \emph{in its electronic ground state}. Although the modified mass $M'$ is not given by the sum of the rest masses of the atomic constituents it is nevertheless invariant under any changes of the internal atomic energy such that we can set $M'=M$ for all practical purposes. This allows us to continue using a notation using energy differences~$\hbar \omega_A$ in~$\mathrm{H}_\text{A}$.

When describing an atom interacting with a laser field of frequency $\omega_L$ it is common to apply a unitary transformation $U=\exp[i (\omega_L \sigma_z + \omega_A \id)t/2]$ such that, for example, $U\ketbra{e}{g} U^\dagger = \ketbra{e}{g} \exp(i \omega_L t)$. The kinetic and atomic Hamiltonian then transform to
\begin{equation}
	\frac{\bP^2}{2M}\left( 1 - \frac{\mathrm{H}_\text{A}}{Mc^2}\right) + \mathrm{H}_\text{A} + i\hbar \dot{U} U^\dagger 
		= \frac{\bP^2}{2 M'}\left( 1 - \frac{\hbar \omega_A}{2 M' c^2} \sigma_z\right) + \frac{\hbar (\omega_A-\omega_L)}{2} \sigma_z - \frac{\hbar \omega_A}{2} \id
		\,.
\end{equation}
The final term can again be dropped without changing the resulting dynamics. One should thus take note that the unitary transformation used, for example, in the rotating wave approximation modifies the effective free atomic Hamiltonian, but not the one appearing in the kinetic term describing the central motion.

\section{Examples}\label{sec:_Examples}
In the following three examples we shall briefly examine the effect of the modified Hamiltonian~\eqref{eq:_Hamiltonian_final}. The first example of a moving atom undergoing spontaneous decay was the trigger for this work. Later we consider the role of this modification for atomic transition rates and Rabi oscillations. 

\subsection{Decaying atom}
The example of a moving, decaying atom was the initial motivation for this work~\cite{sonnleitner2017will}. Using the Hamiltonian from equ.~\eqref{eq:_Hamiltonian_final}, we see that the equations for the central motion in the Heisenberg picture are
\begin{subequations}\label{eq:_eqmotion}
\begin{align}
	\dot\bP &= \tfrac{i}{\hbar} [\mathrm{H}, \bP] = \tfrac{i}{\hbar} [\mathrm{H}_\text{AL}, \bP]
	\,,	\\
	M \dbQ &= \bP\left( 1 - \mathrm{H}_\text{A}/(Mc^2)\right) + \bd \times \bB(\bQ)
	\,,\\
	M \ddot{\bQ} &=\dot\bP\left( 1 - \mathrm{H}_\text{A}/(Mc^2)\right) - \tfrac{i}{\hbar}\bP [\mathrm{H}, \mathrm{H}_\text{A}]/(Mc^2) + \tfrac{i}{\hbar} [\mathrm{H}, \bd \times \bB(\bQ)]
	\,.
\end{align}	
\end{subequations}
In~\cite{sonnleitner2017will} we showed that an initially excited two-level atom $\mathrm{H}_\text{A}=\hbar\omega_A \ketbra{e}{e}$ decaying at a rate~$\Gamma$ will change its internal energy
as $\tder{t}{} \ew{\mathrm{H}_\text{A}} = -\Gamma \hbar \omega_A$ while the average momentum changes as $\tder{t}{}\ew{\bP} = - \Gamma \hbar \omega_A \ew{\bP}/(Mc^2)$ and $\tder{t}{}\ew{\bd \times \bB}=0$. 
As $\bP$ and $\mathrm{H}_\text{A}$ commute we can write for the expectation values 
\begin{equation}
	M \tder{t}{2}\ew{\bQ} = \tder{t}{}\ew{\bP}\left( 1 - \ew{\mathrm{H}_\text{A}}/(Mc^2)\right) - \ew{\bP} \tder{t}{} \ew{\mathrm{H}_\text{A}}/(Mc^2) + \tder{t}{}\ew{\bd \times \bB}
	\,,
\end{equation}
such that $M \tder{t}{2}\ew{\bQ} \approx 0$ as it should be. Without the correction term $\sim \mathrm{H}_\text{A}/(Mc^2)$ we would have had $M \ew{\ddot\bQ} = \ew{\dot\bP}$ with the misleading result that a decaying, moving atom feels an average deceleration during the emission process. This is corrected by the use of the Hamiltonian~\eqref{eq:_Hamiltonian_final}, which correctly distinguishes between a change in momentum due to acceleration and a force due to a change of internal energy.

\subsection{Modification of internal atomic dynamics}
The dynamics of some operator describing the internal states of the atom, for example $\sigma_+ = \ketbra{e}{g}$, are given by $\dot{\sigma}_+ = \tfrac{i}{\hbar} [\mathrm{H}, \sigma_+]$ with
\begin{align}
	[\mathrm{H}, \sigma_+] &= \left[\mathrm{H}_\text{A} \left( 1-\tfrac{\bP^2}{2M^2c^2}\right) + \mathrm{H}_\text{AL}, \sigma_+ \right]
		\approx \left( 1-\tfrac{\bP^2}{2M^2c^2}\right)\left[\mathrm{H}_\text{A} + H_\text{AL}, \sigma_+ \right]
		\notag \\
	&\approx \tfrac{1}{\gamma}\left[\mathrm{H}_\text{A} + \mathrm{H}_\text{AL}, \sigma_+ \right]\,,
\end{align}
with the Lorentz factor $\gamma \approx 1 + \bP^2/(2M^2c^2)$. Thus we see that the Heisenberg equation of motion is actually with respect to proper time, \ie
\begin{equation}
	\tder{\tau}{} \sigma_+ = \gamma \tder{t}{} \sigma_+ = \tfrac{i}{\hbar}\left[\mathrm{H}_\text{A} + \mathrm{H}_\text{AL}, \sigma_+ \right]
	\,.
\end{equation}
This is consistent with the equations of motion for the two charges in the Lagrangian~\eqref{eq:_Lagrangian_final}, which are also the Lorentz force in the proper time.\footnote{Note that the Lorentz factor did not appear in the equations of motion~\eqref{eq:_eqmotion} because we dropped the term~$\bP^4/(8M^3c^2)$}

\subsection{Photon absorption and Rabi oscillations}
Although the motivation for this work was an atom undergoing spontaneous emission, it is clear that also absorption or stimulated emission will change the atom's mass-energy in the manner described above. A careful analysis of these effects should therefore also reveal a ``friction-like'' force which can be correctly interpreted using a Hamiltonian of the type given in equ.~\eqref{eq:_Hamiltonian_final}.

\subsubsection{Fermi's argument for resonant photon absorption}
First, let us briefly review Fermi's classical argument for a moving atom absorbing a photon~\cite{fermi1932quantum,barnett2010recoil,barnett2018vacuumContPhys}. Let an atom move in the $+z$ direction with an initial momentum $p=M v$ encounter a photon of frequency $\omega_L$ propagating in the opposite direction. After absorption the atom will be in an excited state of energy $\hbar \omega_A$ and carry momentum $p^\ast$. Energy and momentum conservation then require that
\begin{subequations}
\begin{align}
	\frac{p^2}{2 M} + \hbar \omega_L &= \frac{{p^\ast}^2}{2 M^\ast} + \hbar \omega_A
	\label{eq:_photon_absorption_energy_conservation}
\,,\\
	p - \hbar \omega_L/c & = p^\ast
	\label{eq:_photon_absorption_momentum_conservation}
\,,	
\end{align}
\end{subequations}
where terms with a star are quantities referring to the excited atom. From this it is easy to obtain the resonance condition including the Doppler and recoil shift
\begin{align}
	\hbar \omega_L\left(1 + \frac{p}{M^\ast c} - \frac{\hbar \omega_L}{2 M^\ast c^2} \right) &= \hbar \omega_A	+ \frac{p^2}{2M}\left( \frac{M^\ast}{M} - 1\right) \,,
	\notag \\
	\omega_L\left(1 + \frac{p}{M c} - \frac{\hbar \omega_L}{2 M c^2} \right) &\approx \omega_A
	\label{eq:_photon_absorption_resonance_condition}
	\,,
\end{align}
where we used $M^\ast\approx M$. The last line can also be written as $\omega_L \approx \omega_A\left(1 - \frac{p}{Mc} + \frac{\hbar \omega_A}{2 M c^2} \right)$. 

In a reference frame where the atom is initially at rest the laser frequency is given by $\omega_L' = \omega_L[1 + p/(Mc)]$ and it is resonant with the atom if $\omega_L' = \omega_A [1 + \hbar \omega_A/(2 M c^2)]$. In this frame we see that the change in the atom's momentum is
\begin{subequations}
\begin{equation}
	(p^\ast)' - 0 = - \hbar \omega_L'/c = -\frac{\hbar \omega_A}{c} \left(1 + \frac{\hbar \omega_A}{2 M c^2}\right)
	\,.
\end{equation}
In comparison, an observer in a frame where the atom is initially moving with momentum $p$ finds
\begin{equation}\label{eq:_photon_absorption_momentum_change}
	p^\ast - p = - \frac{\hbar\omega_A}{c} \left(1 + \frac{\hbar \omega_A}{2 M c^2} \right) + p \frac{\hbar \omega_A}{M c^2}
\,.	
\end{equation}	
\end{subequations}
Thus, we again find a situation in which observers in different reference frames see a different change in momentum and again this change is proportional to the initial momentum multiplied by the change in internal energy relative to the atomic mass.\footnote{Note that here the atomic energy increases during the absorption process, hence the different sign of the ``friction'' term.}

As before, this is resolved by accounting for a change in mass-energy by setting $p^\ast = M^\ast v^\ast$ such that equ.~\eqref{eq:_photon_absorption_momentum_conservation} gives $v^\ast = \left(M v - \hbar \omega_L/c \right)/M^\ast$. The actual change in velocity due to the absorption of a resonant photon is then given by
\begin{align}
	 v^\ast - v &=  v \left(\frac{M}{M^\ast} - 1\right) - \frac{\hbar \omega_L}{M c}\frac{M}{M^\ast} 
	 \notag \\
		&\approx v \left(\frac{M}{M^\ast} - 1\right) - \frac{\hbar\omega_A}{M c} \left(\frac{M}{M^\ast} + \frac{\hbar \omega_A}{2 M c^2} \right) + v \frac{\hbar \omega_A}{M c^2}
	\notag \\
		&\approx - \frac{\hbar\omega_A}{M c} \left(1 - \frac{\hbar \omega_A}{2 M c^2} \right)
		\,. \label{eq:_photon_absorption_acceleration}
\end{align}
For the final result we used $M^\ast = M + \hbar \omega_A/c^2$~\cite{einstein1905tragheit}. 

Before we consider the corresponding quantum case let us highlight a few details of the calculation so far: First we should note that the introduction of a modified mass $M^\ast$ was not necessary to obtain the resonance condition in equ.~\eqref{eq:_photon_absorption_resonance_condition} and indeed usual derivations of this condition do not consider a possible change in mass-energy~\cite{fermi1932quantum,barnett2010recoil}. However, setting $M^\ast = M + \hbar \omega_A/c^2$ is \emph{required} to obtain an acceleration which is independent of the original velocity as one would expect from the case for an atom initially at rest.

Second, we would like to clarify that the result of equ.~\eqref{eq:_photon_absorption_acceleration} does, of course, not contradict Doppler cooling. The probability to absorb a photon and receive a momentum kick is still proportional to the detuning, which remains sensitive to the atom's initial velocity.

It is interesting to note that the Doppler shift in absorption and emission as well as aberration played a crucial role in Einstein's study of thermal equilibrium and the relationship between the Maxwell-Boltzmann velocity distribution and the Planck spectrum of thermal radiation~\cite{einstein1917quantum}.

\subsubsection{A two-level atom in a laser beam: Rabi oscillations}
The discussion above was not based on any Hamiltonian, but on energy and momentum conservation. Hence, it is a suitable testing ground for our Hamiltonian which should resolve the ambiguity between change in (canonical) momentum and observed acceleration. 

Including the mass-energy correction laid out in~\eqref{eq:_Hamiltonian_final}, the total Hamiltonian for a two-level atom interacting in one dimension with a semi-classical laser field is
\begin{subequations}
\begin{equation}
	\mathrm{H} = \frac{P^2}{2M} \left( 1 - \frac{\hbar \omega_A}{M c^2} \ketbra{e}{e} \right) + \mathrm{H}_\text{A} + \mathrm{H}_\text{AL}
\,,
\end{equation}
where the mass~$M$ refers to the mass of the atom in the ground state. The field shall be propagating in the $-z$-direction such that we find after a suitable unitary transformation,
\begin{equation}
	\mathrm{H}_\text{A} + \mathrm{H}_\text{AL} = \frac{\hbar \delta}{2} \sigma_z + \frac{\hbar \Omega_L}{2} \left[ \sigma_+ \left(1 + \frac{\mathrm{P} + \hbar k/2}{Mc}\right) e^{-i k \mathrm{Z}} + \Hc \right]
\,,	
\end{equation}
\end{subequations}
where $\delta = \omega_A - \omega_L$ is the detuning, $\hbar \Omega_L$ gives the coupling strength between the dipole and the electric field, $k=\omega_L/c$ is the wavenumber and $[\mathrm{Z},\mathrm{P}]=i\hbar$. The term $\sim (\mathrm{P} + \hbar k/2)/(Mc)$ is due to the Röntgen interaction~\cite{sonnleitner2017rontgen}. For the two-level atom with states $\ket{g}$ and $\ket{e}$ we have $\sigma_z = \ketbra{e}{e}-\ketbra{g}{g}$, $\sigma_+ = \ketbra{e}{g}$ and $(\sigma_+)^\dagger = \sigma_-$.

Solving the Schrödinger equation for a general atomic state $\ket{\psi(t)} = \int dp ( c_g(t,p)\ket{g,p} + c_e(t,p)\ket{e,p})$ with initial condition $c_g(0,p)=c_g^0(p)$ and $c_e(0,p)=0$ we get
\begin{subequations}
\begin{align}
	c_g(t, p) &= c_g^0(p) e^{-i \Phi(p) t} \left[ \cos\left(\Omega_R(p) t/2\right) + i \frac{\Delta(p)}{\Omega_R(p)} \sin\left(\Omega_R(p) t/2\right) \right]
\,,\\
	c_e(t, p-\hbar k) &= - i c_g^0(p) \frac{\Omega_L}{\Omega_R(p)} \left(1 + \frac{p - \hbar k/2}{Mc}\right) e^{-i \Phi(p) t} \sin\left(\Omega_R(p) t/2\right) 
\,.
\end{align}
Here
\begin{align}
	\Phi(p) &= \frac{p^2 + (p-\hbar k)^2}{4 \hbar M}
\,, \\
	\Delta(p) &= \omega_A - \omega_L\left(1 + \frac{p-\hbar k/2}{Mc}\right)
\,, \\
	\Omega_R^2(p) &= \Omega_L^2 \left(1 + 2 \tfrac{p - \hbar k/2}{Mc}\right) + \Delta^2(p)
\,,	
\end{align}
\end{subequations}
give the dispersion of the wave packet, the detuning between the moving atom and laser field including the recoil shift and the Rabi frequency, respectively. As one would expect, the ground-state amplitude for momentum~$p$ is coupled to the excited-state amplitude for momentum $p-\hbar k$ and $\abs{c_g(t, p)}^2 + \abs{c_e(t, p-\hbar k)}^2 = \abs{c_g^0(p)}^2$. For an optimally chosen laser frequency we find $\Delta(p)=0$ and the atom oscillates between the ground and excited state at a rate given by the Rabi frequency.

Let $p_0 = \int dp\, p \abs{c_g^0(t)}^2$ denote the momentum expectation value for the initial wave packet which is well localised in momentum space such that $\int dp\, f(p) \abs{c_g^0(t)}^2\approx f(p_0)$ for any slowly varying function $f$. Setting $C_e(t) := \int dp\, \abs{c_e(t,p)}^2$ we then find that the momentum expectation value changes as
\begin{subequations}
\begin{equation}
	\ew{\dot{\mathrm{P}}} = - \hbar k \dot{C}_e(t)
	\,,
\end{equation}
while the actual acceleration is given by
\begin{equation}
	M \ew{\ddot{\mathrm{Z}}} = \ew{\dot{\mathrm{P}}} - \frac{\hbar \omega_A}{c} \frac{p_0 - \hbar k}{M c} \dot{C}_e(t) + F_\text{R}
	\,,
\end{equation}
\end{subequations}
where $F_\text{R}$ is a force given by the Röntgen term,
\begin{align}
	F_\text{R} &= \frac{\hbar \Omega_L}{c} \der{t}{} \int dp \re\left[ c_g(t,p) c_e^\ast(t,p-\hbar k)\right] 
	\notag \\
		&= - \frac{\hbar \Omega_L^2}{c} \frac{\Delta(p_0)}{2 \Omega_R(p_0)} \left(1 + \frac{p_0 - \hbar k/2}{Mc}\right) \sin\left(\Omega_R(p_0) t\right)
		\,.
\end{align}
Choosing $\omega_L = \omega_A\left(1 - \frac{p_0 -\hbar k/2}{Mc} \right)$, such that $\Delta(p_0)=0$, we obtain
\begin{subequations}
\begin{equation}
	\ew{\dot{\mathrm{P}}} = - \frac{\hbar \omega_A}{c} \left(1 - \frac{p_0 - \hbar k/2}{Mc} \right) \dot{C}_e(t)
	\,,
\end{equation}
while the actual acceleration is given by
\begin{align}
	\ew{\ddot{\mathrm{Z}}} & = - \frac{\hbar \omega_A}{M c} \left(1 - \frac{p_0 - \hbar k/2}{Mc} \right) \dot{C}_e(t) - \frac{\hbar \omega_A}{M c} \left(\frac{p_0 -\hbar k}{M c} \right)\dot{C}_e(t)
	\notag \\
	 & \approx - \frac{\hbar \omega_A}{M c} \left(1 - \frac{\hbar \omega_A}{2 M c^2} \right) \dot{C}_e(t)
	\,,
\end{align}
\end{subequations}
which is consistent with what we had in equ.~\eqref{eq:_photon_absorption_acceleration} and is independent of the initial momentum $p_0$.

\section{Conclusion}

The usual Hamiltonian describing the non-relativistic interaction between (moving) atoms and external fields is not able to distinguish between the atom's mass and its mass-energy. This is because the motion of atoms is usually described in terms of Galilean physics while electromagnetic fields inherently follow Lorentzian symmetry. This hotchpotch leads to the surprising effect that the concept of mass-energy enters the atomic equations of motion through the back-door of first order the Doppler effect and aberration~\cite{sonnleitner2017will,barnett2018vacuum}.

Motivated by this discrepancy, we re-derived the Hamiltonian for atom-light interaction including some next-order relativistic effects. Starting from Darwin-Lagrangian we first derived a corresponding minimal-coupling Hamiltonian. After a PZW-transformation to a multi-mode Hamiltonian in center-of-mass coordinates, a further canonical transformation gave the final Hamiltonian~\eqref{eq:_Hamiltonian_final}. The essential difference between this Hamiltonian and the usual description is that the kinetic energy term changes as $\bP^2/(2M) \rightarrow \bP^2/(2M)\big(1-\mathrm{H}_\text{A}/(Mc^2)\big)$.

The examples given in section~\ref{sec:_Examples} illustrate that this allows a clear distinction between forces connected to an actual change in the motion of the atom and those that arise due to changes in internal energy. Such an ambiguity can arise whenever mechanical interactions between atoms and light are calculated to the level of a single photon recoil.

Although we only used a very simple atomic model of two opposite charges, there is no reason to assume why more elaborate models of atoms should give very different results, provided the hierarchy of energy scales given in section~\ref{sec:_energy_hierarchy} is preserved.

Finally, we note that the results presented here are reminiscent of those obtained for the motion of a spin-half relativistic dipole, such as a neutron, moving in an external field~\cite{simonovits1993behavior}. Indeed we could have based our analysis on the Dirac equation for a neutron~\cite{bjorken1964relativistic} and extended this, by analogy, to an atomic dipole. We shall present this complimentary derivation elsewhere.

\section*{Acknowledgements}
We are grateful to Simon Horsley, Robert Cameron, Helmut Ritsch and Nils Trautmann for valuable discussion and Carsten Henkel for making us aware of some references. This work was supported by the Austrian Science Fund (FWF) (Grant No. J 3703) and a Royal Society Research Professorship (Grant No. RP150122).

\appendix

\section{Calculation of fields generated by the charges}\label{appendix:_internal fields}
The internal potentials can be calculated from Maxwell's equations in Coulomb gauge, $\nabla^2\phi = -\rho/\epsilon_0$ and $\big(\nabla^2 - \tfrac{1}{c^2} \tpder{t}{2}\big) \bcA^\perp = -\mu_0 \bj^\perp$, with $\bj^\perp = \bj - \bj^\parallel = \bj - \varepsilon_0 \nabla \tpder{t}{}\phi$. These can be solved using the Green's function for the Poisson and wave equation, respectively~\cite{jackson1975classical},
\begin{align}
	\phi(\bx,t) &= \frac{1}{4 \pi \varepsilon_0} \int d^3\bx' \frac{\rho(\bx',t)}{\abs{\bx-\bx'}}
	\label{eq:_phi_internal}	
	\, \\
	\bcA^\perp(\bx,t) &= \frac{\mu_0}{4 \pi} \int d^3\bx'\frac{1}{\abs{\bx-\bx'}} \left[ \bj(\bx',t) - \epsilon_0 \nabla_{\bx'} \tpder{t}{}\phi(\bx',t) \right]_\text{ret}
		\notag \\
		&= \frac{\mu_0}{4 \pi} \int d^3\bx'\frac{1}{\abs{\bx-\bx'}} \left[ \bj(\bx',t) + \frac{\nabla_{\bx'}}{4 \pi} \int d^3\bx'' \frac{\nabla_{\bx''} \cdot \bj(\bx'',t)}{\abs{\bx'-\bx''}} \right]_\text{ret}
	\,.	
\end{align}
Here we used the continuity equation $\nabla \cdot \bj + \partial_t \rho = 0$ to obtain the last line and $\nabla_{\bx'}$ indicates derivation with respect to primed coordinates. The brackets $[\dots]_\text{ret}$ indicate that an expression is evaluated at the retarded time $t' = t-\abs{\bx-\bx'}/c$, but as $\bcA^\perp$ appears in the Lagrangian only together with $\bj$, including retardation would lead to terms in third order of velocity and go beyond our approximation. Using partial integration we obtain the vector potential generated by the moving charges~\cite{jackson1975classical},
\begin{align}
	\bcA^\perp(\bx,t) &\simeq 
		\frac{\mu_0}{4 \pi} \int d^3\bx' \frac{\bj(\bx',t)}{\abs{\bx-\bx'}} + \frac{\mu_0}{(4 \pi)^2} \int d^3\bx' \int d^3\bx'' \frac{\bx-\bx'}{\abs{\bx-\bx'}^3}  \frac{\bj(\bx'',t)\cdot(\bx'-\bx'')}{\abs{\bx'-\bx''}^3} 
	\notag \\
		&= 
		\frac{\mu_0}{8 \pi} \sum_{i=1,2} e_i \left( \frac{\dbr_i}{\abs{\bx-\br_i}} + \frac{(\bx-\br_i) [\dbr_i\cdot(\bx-\br_i)]}{\abs{\bx-\br_i}^3} \right)
		\,.
	\label{eq:_Aperp_internal}	
\end{align}

\section{Negligible terms in the Lagrangian~\texorpdfstring{\eqref{eq:_L_intermediate}}{}}\label{appendix:_evil_terms}

The Lagrangian given in equ.~\eqref{eq:_L_intermediate} contains two unfamiliar terms which shall be examined here. 

The first is $\frac{\varepsilon_0}{4} \int d^3\bx \tpder{t}{2} (\bcA^\perp)^2$, the second time-derivative of the internal vector potential. From equ.~\eqref{eq:_Aperp_internal} we see that $\varepsilon_0(\bcA)^2$ is proportional to terms of second order in the particle velocities divided by the speed of light~$c$. The additional time-derivatives of $\partial_t^2 (\bcA^\perp)^2$ will then give terms $\sim \abs{\dbr_i}^2\abs{\dbr_j}^2/c^4$ as well as terms containing accelerations $\abs{\ddot{\br}_i} \abs{\dbr_j}^2/c^4$ and even $\abs{\dddot{\br}_i} \abs{\dbr_j}/c^4$, ${i,j=1,2}$. We can assume that the dominant force in our setup is the electrostatic attraction between the particles, so that
\begin{equation}
	\frac{\abs{\ddot{\br}_{i}} \abs{\dbr_j}^2}{c^4} \propto \frac{1}{4 \pi \varepsilon_0} \frac{e_1 e_2}{\abs{\br_1-\br_2}} \frac{\abs{\dbr_j}^2}{m_i c^4}
	\,.
\end{equation}
These terms are thus proportional to the electrostatic energy of the atom divided by $m_i c^2$ times $\frac{\abs{\dbr_j}^2}{c^2}$, which goes beyond our level of approximation. Other works deriving the Darwin Lagrangian have used similar arguments~\cite{breit1932dirac,close1970relativistic}.

The second term under consideration here is the final term of equ.~\eqref{eq:_L_intermediate}, 
\begin{equation}\label{eq:_self-field}
	\varepsilon_0 \int d^3\bx \big((\partial_t \bcA^\perp)\cdot(\partial_t \bA^\perp) - c^2 (\nabla \times \bcA^\perp)\cdot(\nabla\times\bA^\perp) \big)
	\,.
\end{equation}
This is a cross term between the transverse electric and magnetic fields generated by the moving charges and the external fields. Terms like this are not specific to our problem, they appear whenever the back-action of fields generated by a (moving) charged particle on itself are considered. This term is thus connected to radiation reaction, electromagnetic masses and similar problems~\cite{jackson1975classical,rohrlich1964solution,rohrlich2007classical}. As we are also ignoring other formally infinite self-action terms appearing in $\bj\cdot\bcA - \rho \phi$ we can also drop the term~\eqref{eq:_self-field} following the same rationale.

\section{Proof that one cannot construct canonical variables using the center of energy}\label{appendix:_proof_no_CoE}

Throughout this work we have made the claim that it is impossible to use the center of energy as a central coordinate for our problem to two charged particles. In equ.~\eqref{eq:_coe_compared_to_Q} we gave the center of energy in terms of the canonical quantities $\bQ, \bq, \bP, \bp$ and found
\begin{equation}\label{eq:_coe-Q}
	\bfR - \bQ = \frac{1}{4 M^2 c^2}\left[ \bq(\bP\cdot\bp) + (\bp\cdot\bP)\bq - \bp(\bP\cdot\bq) - (\bq\cdot\bP)\bp \right] \,.
\end{equation}
$\bQ, \bq$ are canonical coordinates with their respective momenta $\bP,\bp$ and a straightforward (but lengthy) calculation confirms that $[\bQ_k, \bP_l] = [\bq_k, \bp_l] = i\hbar \delta_{kl}$ and $[\bQ_k, \bq_l] = [\bQ_k, \bp_l] = [\bP_k, \bp_l] = [\bP_k, \bq_l] = 0$. Using the relationship between $\bQ$ and $\bfR$ we shall now try to construct a set of canonical coordinates $(\bfR, \bP_{\bfR}; \bfr, \bp_{\bfr})$ which preserve these commutation relations.

The central momentum $\bP=\bp_1+\bp_2$ already is the total momentum of our two particles. If $\bfR$ is to be a central coordinate, then we need $\bP_{\bfR}=\bP$, and this also fulfils $[\bfR_k,\bP_l]=[\bQ_k,\bP_l]$.

It is also reasonable to demand that the new set of coordinates converges towards the center of mass coordinates in the lowest approximation, especially $\lim_{c\rightarrow\infty} \bfr = \br = \br_1-\br_2$. We can therefore say that $\bfr - \bq$ is ``small'' just as $\bfR - \bQ$ above such that $(\bfR - \bQ)_k (\bfr - \bq)_l \approx 0$. Using this reasoning, we can write
\begin{align}
	[\bfR_k, \bfr_l] &= [\bQ_k + (\bfR - \bQ)_k, \bq_l + (\bfr - \bq)_l] \notag \\
		&\approx [\bQ_k, \bq_l] + [\bQ_k, (\bfr - \bq)_l] + [(\bfR - \bQ)_k, \bq_l]
		\,,
\end{align}
such that, using $[\bQ_k, \bq_l] = 0$ and demanding $[\bfR_k, \bfr_l] \stackrel{!}=0$, we find
\begin{equation}\label{eq:_coe_not_canonical_interm1}
	[\bQ_k, (\bfr - \bq)_l] \stackrel{!}{=} - [(\bfR - \bQ)_k, \bq_l]
	\,.
\end{equation}
With equ.~\eqref{eq:_coe-Q} we find
\begin{equation}\label{eq:_coe_not_canonical_interm2}
	[(\bfR - \bQ)_k, \bq_l] = - \frac{i \hbar}{2M^2c^2} \left( \bq_k \bP_l - \delta_{lk} (\bq\cdot\bP) \right)
		= - \frac{i \hbar}{2M^2c^2} \sum_n \left( \delta_{ln} \bq_k - \delta_{lk} \bq_n \right) \bP_n
	\,.
\end{equation}
As $[\bQ_k, (\bfr - \bq)_l] = i \hbar \tpder{\bP_k}{} (\bfr - \bq)_l$ we see that $(\bfr - \bq)$ has to be of second order in $\bP$. We thus set 
\begin{equation}
	(\bfr - \bq)_l = \sum_{mn} \alpha_{lmn}(\bq) \bP_m \bP_n + C_l(\bq,\bp) \,,
\end{equation}
for a set of coefficients $\alpha_{lmn}(\bq)$ and some self-adjoint $C_l(\bq,\bp)$ which commutes with $\bQ_k$. We then find
\begin{equation}\label{eq:_coe_not_canonical_interm3}
	[\bQ_k, (\bfr - \bq)_l] = i \hbar \sum_{mn} \alpha_{lmn}(\bq) \left(\delta_{nk}\bP_m + \delta_{mk} \bP_n\right)
		=  i \hbar \sum_{n} \left( \alpha_{lnk}(\bq) + \alpha_{lkn}(\bq) \right)\bP_n
	\,.	
\end{equation}
If equ.~\eqref{eq:_coe_not_canonical_interm1} should hold, then we must find that each term in the sums of equ.~\eqref{eq:_coe_not_canonical_interm2} and~\eqref{eq:_coe_not_canonical_interm3}, respectively, are the same, \ie
\begin{equation}
	- \frac{1}{2M^2c^2}\left( \delta_{ln} \bq_k - \delta_{lk} \bq_n \right) \stackrel{!}{=} \left( \alpha_{lnk}(\bq) + \alpha_{lkn}(\bq) \right)
	\,.
\end{equation}
However, we see that the expression on the left-hand side (l.h.s.) is antisymmetric under exchange of $n$ and $k$ while the right-hand side (r.h.s.) is symmetric ($[\text{l.h.s}]_{n,k} = - [\text{l.h.s}]_{k,n}$ while $[\text{r.h.s}]_{n,k} = [\text{r.h.s}]_{k,n}$). This shows that it is impossible to fulfil equ.~\eqref{eq:_coe_not_canonical_interm1} and thus $[\bfR_k, \bfr_l] \neq 0$ for any relative coordinate~$\bfr$.


\end{document}